\documentclass{iaus}
\usepackage{graphicx}

\title[IAU268.~~Light elements in stars with exoplanets] 
{Light elements in stars with exoplanets}

\author[Nuno C. Santos \etal\ ]   
{N.C. Santos$^1$, E. Delgado Mena$^2$, G. Israelian$^2$, J. I. Gonz\'alez-Hern\'andez$^3$, M. C. G\'alvez-Ortiz$^4$, M. Mayor$^5$, S. Udry$^5$, R. Rebolo$^{2,6}$, S. Sousa$^1$ \and S. Randich$^7$}

\affiliation{
$^1$Centro de Astrof\'isica, Universidade do Porto, Rua das Estrelas, 4150-762 Porto, Portugal. email: {\tt Nuno.Santos@astro.up.pt} \\[\affilskip]
$^2$Instituto de Astrof\'isica de Canarias, E-38200 La Laguna, Tenerife, Spain. \\ [\affilskip]
$^3$Departamento de Astrof\'isica, Facultad de Ciencias F\'isicas, Universidad Complutense de Madrid, E-28040, Spain. \\[\affilskip]
$^4$Centre for Astrophysics Research, Science and Technology Research Institute, University of Hertfordshire, Hatfield AL10 9AB, UK. \\[\affilskip]
$^5$Observatoire de Gen\` eve, 51 ch. des Maillettes, CH-1290 Sauverny, Switzerland. \\[\affilskip]
$^6$Consejo Superior de Investigaciones Cient\'{\i}ficas, E-28006, Madrid, Spain.\\[\affilskip]
$^7$INAF/Osservatorio Astrofisico di Arcetri, Largo Enrico Fermi 5, I-50125 Firenze, Italia.}

\pubyear{2010}
\volume{268}  
\jname{Light elements in the Universe}
\editors{C. Charbonnel, M. Tosi, F. Primas \& C. Chiappini, eds.}
\begin{document}

\maketitle

\begin{abstract}
It is well known that stars orbited by giant planets have higher abundances of heavy elements when compared with
average field dwarfs. A number of studies have also addressed the possibility that light element abundances are
different in these stars. In this paper we will review the present status of these studies. The most significant trends will be discussed.
\end{abstract}

\firstsection 
\section{Introduction}

Since the discovery of the first exoplanet around 51\,Peg (\cite{Mayor}) the number of known exoplanets orbiting solar type stars did not stop growing, 
being currently of more than 400 (including more than 30 multi-planetary systems)\footnote{See tables at http://www.exoplanet.eu}. In addition, more than 50 planets are now known to transit their host stars. Finally, in the last few years about 30 planets with masses between 2 and 20 $M_{Earth}$ have been discovered. Present results strongly 
suggest that planets are common around solar-type stars.

\begin{figure}[ht!]
\begin{center}
 \includegraphics[width=6.7cm]{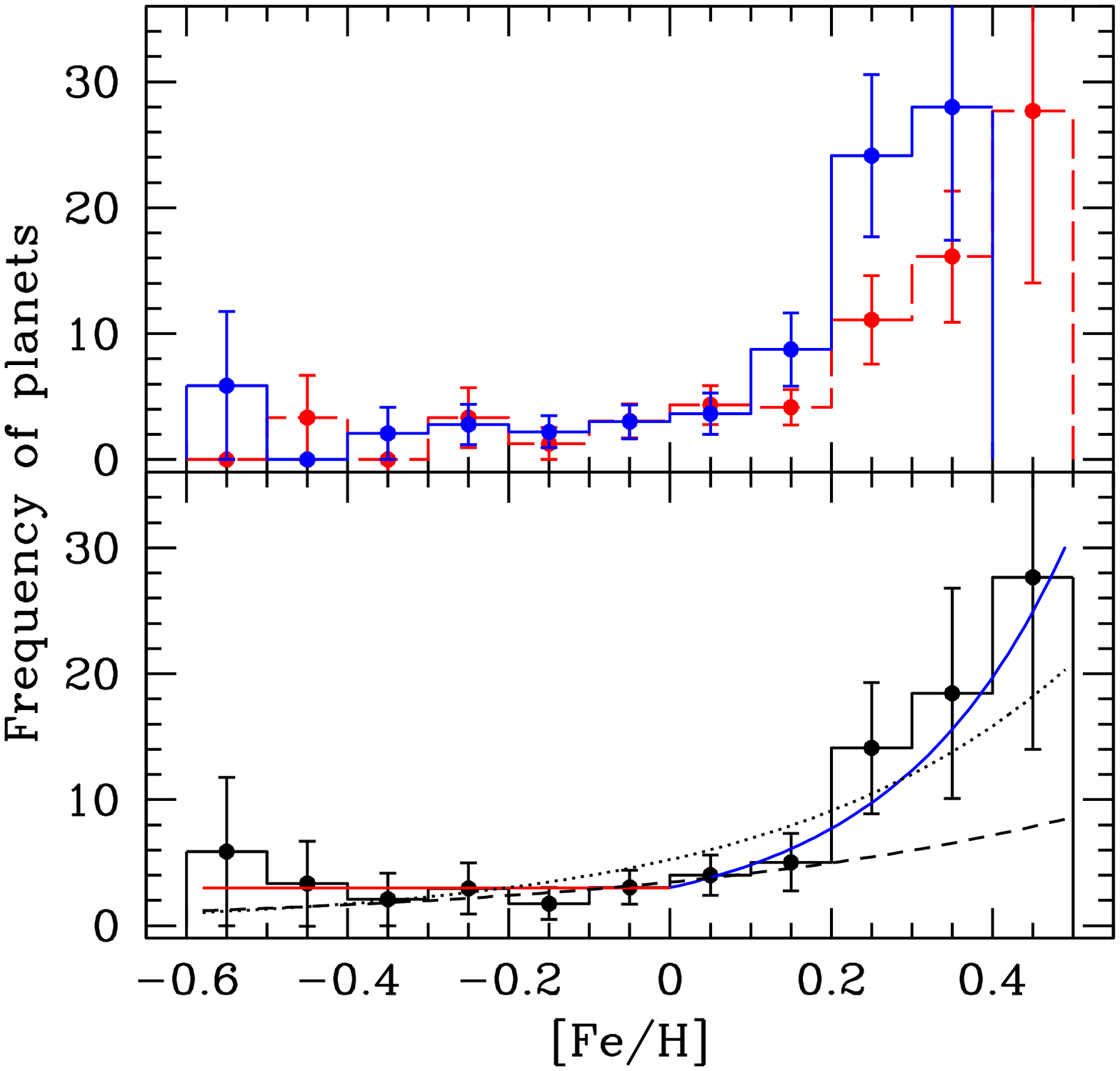} 
 \includegraphics[width=6.7cm]{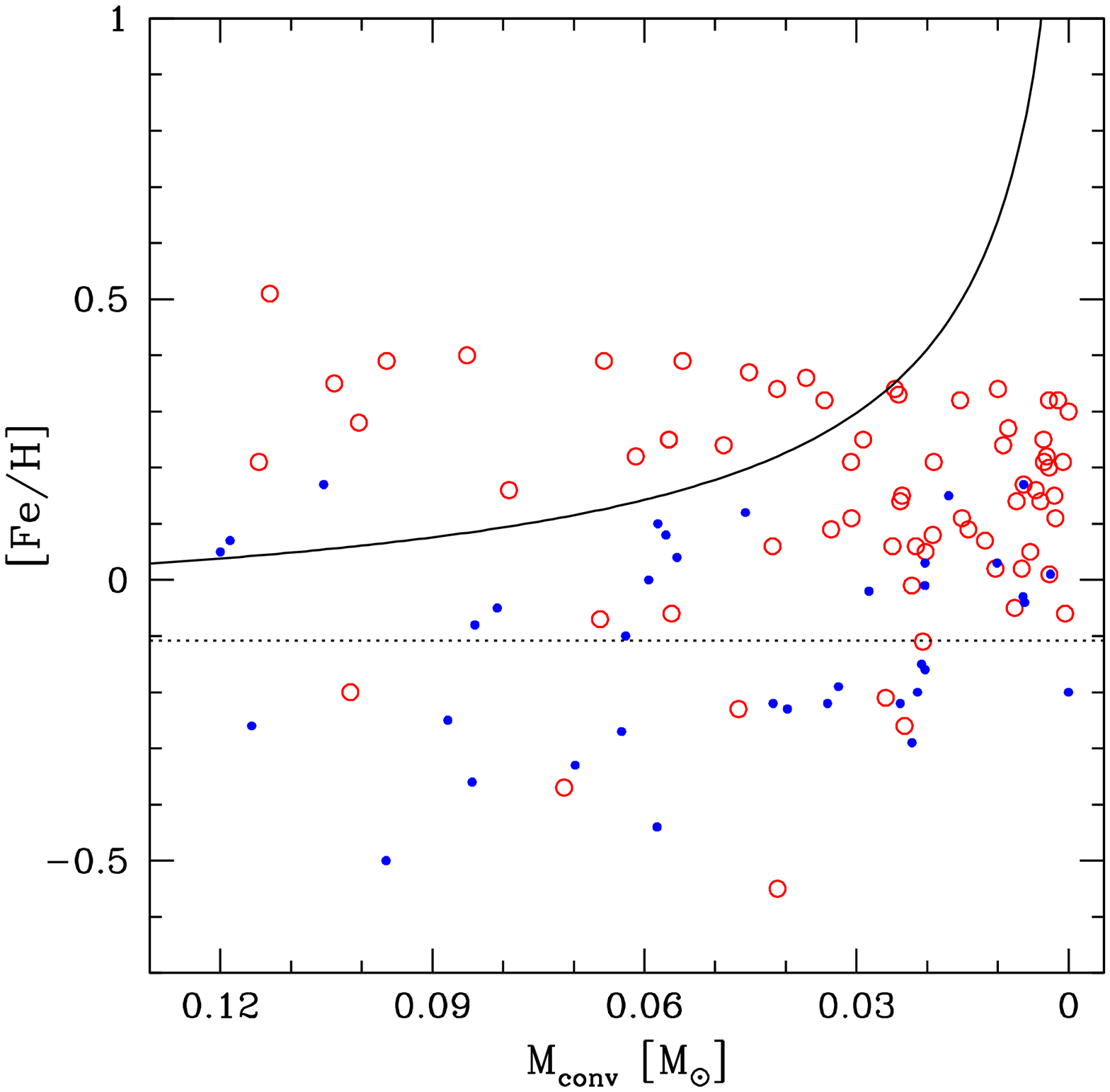} 
 \caption{\textit{Left panel:} Frequency of planet hosts as a function of stellar metallicity. Blue points are from \cite{Santos04a} and red points are from \cite{Fischer}. From Udry \& Santos (2007). \textit{Right panel:} Metallicity as a function of convective envelope mass for stars with planets (open symbols) and field stars (points). The [Fe/H] = constant line represents the mean [Fe/H] for the non-planet hosts stars from Santos \textit{et al.} (2001a). The curved line represents the result of adding 8 earth masses of iron to the convective envelope of stars having an initial metallicity equal to the non-planet hosts mean [Fe/H]. The resulting trend has no relation with the distribution of the stars with planets. From \cite{Santos03}.}
   \label{planets}
\end{center}
\end{figure}

One remarkable characteristic of planet host stars is that they are considerably metal rich when compared with single field dwarfs (\cite{Gonzalez98}; \cite{Santos00}; \cite{Santos04a}; \cite{Fischer}). As we can perfectly see in Fig.\ref{planets}, the probability of finding a giant planet depends strongly on the metallicity of the star. Interestingly, as discussed in Santos \etal\ (2004a), there seems to be two regimes in this distribution. For super solar metallicities the presence of planets is a steep rising function of metallicity, while for metal poor stars there is no significant dependence with metallicity. 

Two main explanations have been suggested to explain the observed metallicity ``excess". In the first one it is considered that its origin is primordial, so the more metals you have in the proto-planetary disk, the higher should be the probability of forming a planet. Alternatively, this excess could be produced by accretion of rocky material by the star sometime after it reached the main-sequence.

In the right panel of Fig.\ref{planets} metallicity is shown as a function of the stellar convective envelope mass. If pollution were the main responsible for the enhanced metallicity of planet hosts, we would expect to find higher metallicities as the convective envelope mass decreases. We do not find such a trend. Observations of stars arriving at the main sequence in open clusters do not show this correlation either (\cite{Shen}). Another point against pollution is that we would need too high accretion rates to explain the metallicity observed in K-dwarfs. In addition, transit detections showed that the mass of heavy elements in the planets appears to be correlated with the metallicity of their parent stars (\cite{Guillot06}). The stars that are the most metal-rich host the most metal-rich planets. Finally, a recent work by \cite{Mordasini} finds that the distributions of planetary properties are well reproduced using core-accretion models (see below), which are dependent on dust content of the disk, thus supporting the primordial origin of overmetallicity in stars with planets.

These results have important implications for the models of giant planet formation and evolution.
Two major giant planet formation models have been proposed. The core accretion model (\cite{Pollack}) and the disk instability model (\cite{Boss}). In the former case, planets are formed by the collisional acumulation of planetesimals by a growing solid core, followed by accretion of a gaseous envelope onto the core. In the second one, a gravitationally unstable region in a protoplanetary disk forms self-gravitating clumps of gas and dust (\cite{Boss}). In the core accretion model, planet formation is critically dependent of the dust content of the disk (\cite{Pollack}) while in the disk instability model it is not (\cite{Boss02}). Present observations are thus more compatible with core accretion model, though do not exclude disk instability. 

Although most data suggest that pollution is not the major source of the high metallicity levels in planet host stars, this issue is still not settled.
In contrast with main sequence stars, planet-hosting giants do not show a tendency of being more metal rich. \cite{Pasquini_a} proposed that the lack of a metallicity-planet connection among giant stars is due to pollution of the star while on the main sequence, followed by dillution during the giant phase. Moreover, if hydrogen-poor material is accreted in the early phases of stellar evolution, during planet formation, the metal excess in the convective zone could be diluted, first by dynamical convection, then by thermohaline mixing, on a timescale much shorter than the stellar lifetime (\cite{Vauclair04}).

Although the primordial origin of metallicity seems to be the more compatible explanation, pollution might have been able to alter more or less significantly the global metallicity of stars. In fact, though at low levels, cases of [Fe/H] pollution exist (e.g. \cite{Laws}). In a different context, some Li-rich giants have been found (\cite{Brown}), stars which should have depleted their lithium due to their deep convective envelopes, and that perhaps might have accreted metal-rich material. However, none of the Li-rich giants studied by \cite{Melo} were found to have detectable Be, something incompatible with an engulfment scenario (see also review by V. Smith in this book).

The key to understand these issues may be in the study of light element abundances.  Light element should be particularly sensitive, since they are normally depleted in solar-type stars. If they are present in large quantities, a external origin might be the best explanation, rather than stellar evolution. Furthermore, light elements are important tracers of internal structure and mixing in stars and they can give us information about the rotational history of the star. It is indeed plausible that the planet formation process is able to alter the rotational history of a star, thus inducing changes in the observed abundances of light elements. 

In this paper we will review the major results of 
the study of the light elements Li and Be in stars with planets.

\section{$^{6}$Li: Tracing pollution events.}

The lithium isotope $^{6}$Li is produced by spallation reactions in the insterstellar medium while it is easily destroyed in stellar interiors at a temperature of 2 million K. According to standard models (\cite{Forestini}, \cite{Montalban02}), at a given metallicity there is a mass range, where $^{6}$Li but not $^{7}$Li is being destroyed. These models predict that no $^{6}$Li can survive pre-MS mixing in metal-rich solar-type stars. We note, however, that the mass and the depth of the convection zone also depend on the metal content of the star, and for this reason several old metal-poor stars have preserved a fraction of their initial $^{6}$Li nuclei (\cite{Cayrel}). In any case, these results suggests that any detected $^{6}$Li in a metal-rich solar-type star would most probably be a signal of an external source, that is, accretion of planetary-like material (\cite{Sandquist}).
It is worth mentioning that $^{6}$Li cannot be produced in large quantities in stellar flares, although some is produced (\cite{Ramaty00}). 

\begin{figure}[ht!]
\begin{center}
 \includegraphics[width=6.7cm]{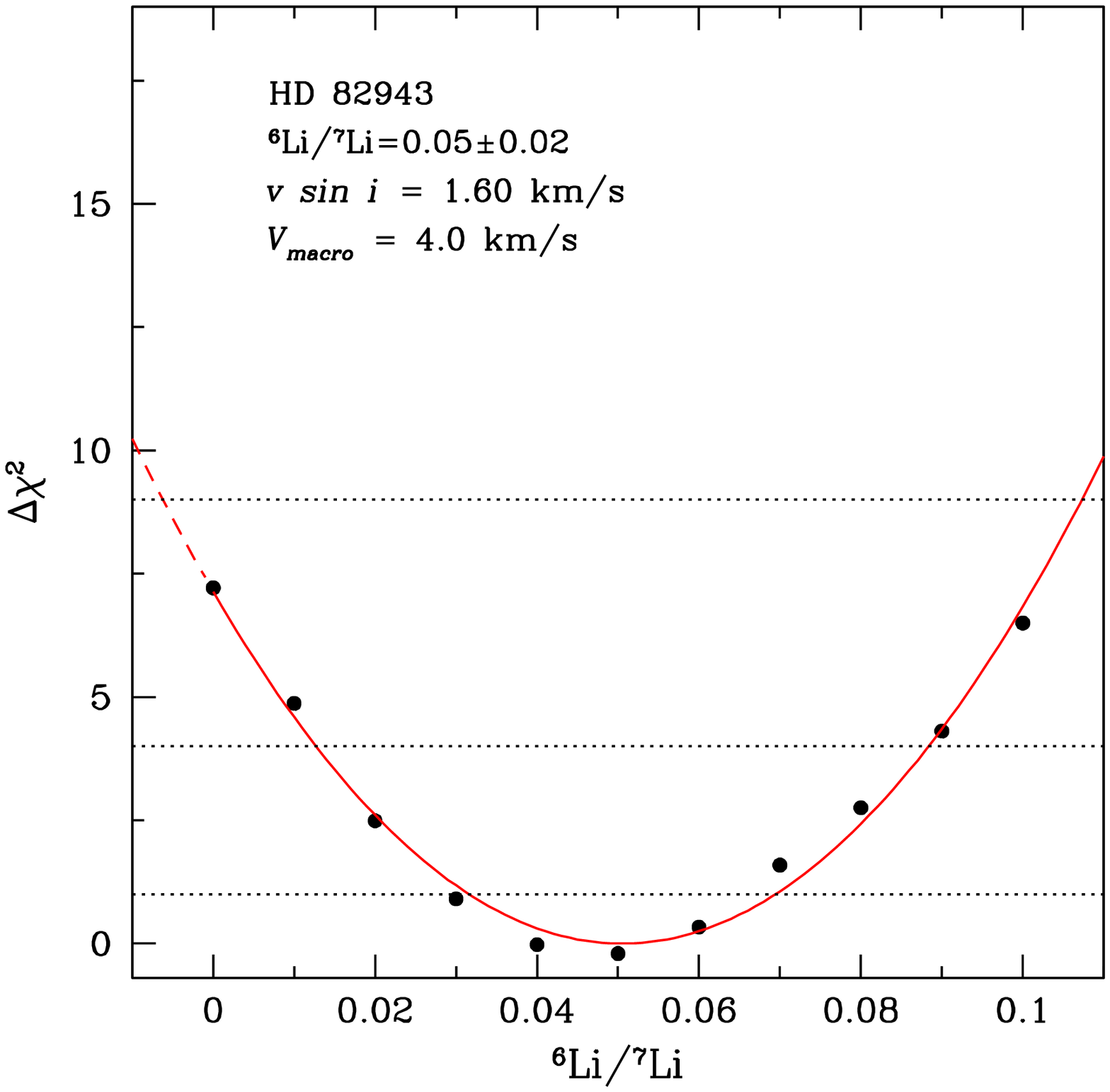} 
 \includegraphics[width=6.7cm]{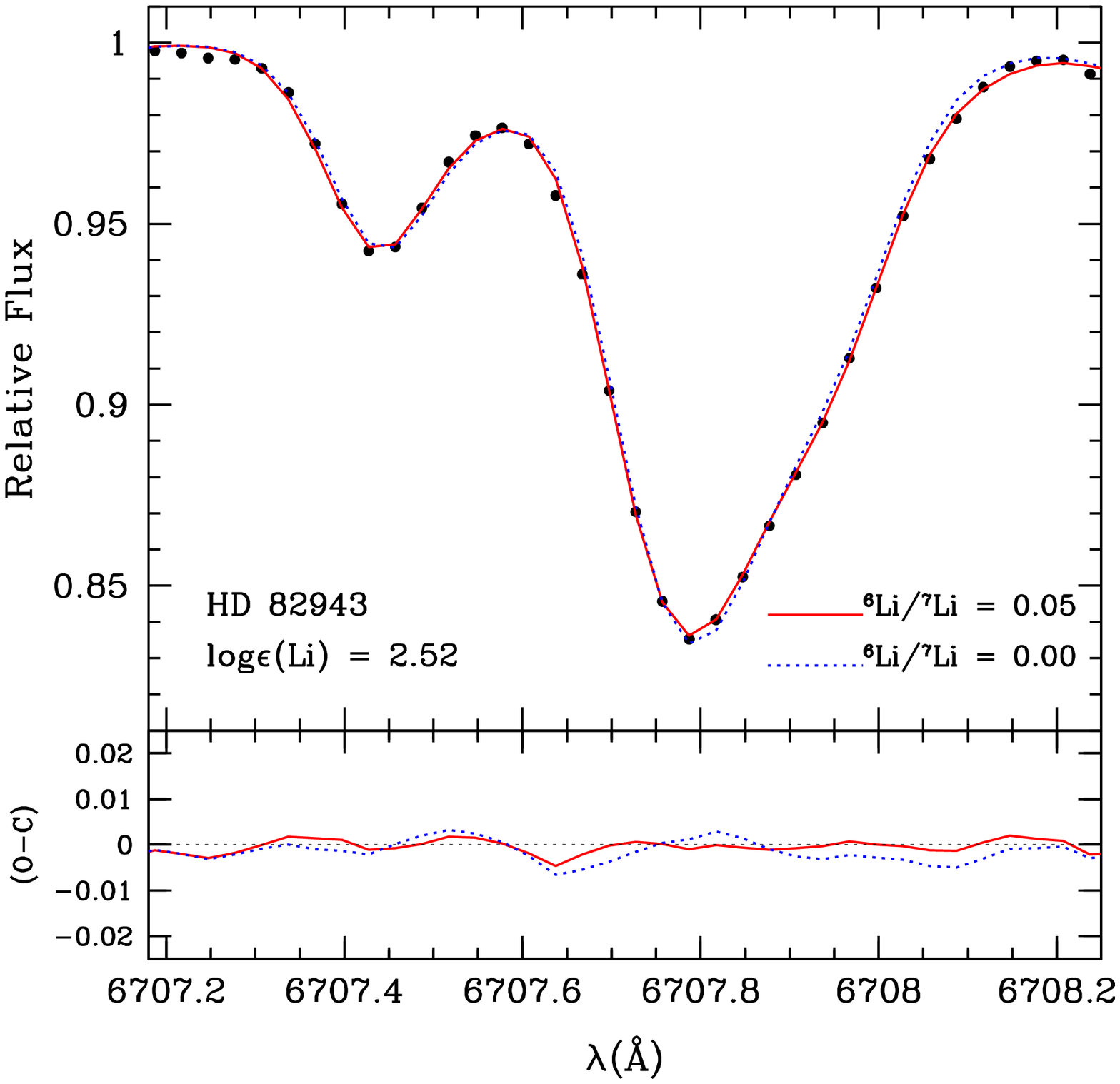} 
 \caption{\textit{Left panel:} Results from the $\chi ^2$ analysis for the $^{6}$Li/$^{7}$Li ratio. $\Delta \chi ^2$= 1, 4, and 9 correspond to 1, 2 and 3$\sigma$ confidence limit (dotted lines). Continuum adjustments were allowed within 0.2$\%$. \textit{Right panel:} Comparison of the observed (filled large dots) and synthetic spectra of HD 82943 corresponding to $^{6}$Li/$^{7}$Li = 0.05 (continuous) and $^{6}$Li/$^{7}$Li = 0 (dotted) isotopic ratios. They correspond to the best fits with f($^{6}$Li) = 0 and a wavelength offset of -0.65 km s$^{-1}$ (small dots) and to the f($^{6}$Li) = 0.05 with a wavelength offset of -0.44 km s$^{-1}$ (continuous line). Fits to the blue wing of the Li profile can be improved if we adopt the wavelengths of CN lines from Brault \& Müller (1975). The CN lines observed in the arc spectrum by these authors appear at 6707.55 Å while Reddy et al. (2002) and Lambert et al. (1993) list them between 6707.464 and 6707.529 Å. The residuals (O-C) of the observations after subtraction of the synthetic spectra are shown. Both plots are from \cite{Israelian03}.}
   \label{li6}
\end{center}
\end{figure}

The first $^{6}$Li detection in a planet host star was reported by Israelian \etal\ (2001, 2003) in HD82943, a star which hosts two planets. This is an old and non active G0 dwarf with an effective temperature of 6010 K, [Fe/H]=$+$0.32 and log N(Li) $\sim$ 2.5. They found an isotopic ratio $^{6}$Li/$^{7}$Li $\sim$ 0.05 (see Fig. \ref{li6}), which may be explained by infall of 1 M$_J$ or equivalent, a value that would not significantly alter the stellar metallicity.

Observations of $^{6}$Li are very difficult because it is a weak component of the much stronger doublet of $^{7}$Li. Moreover, in metal rich stars, blending with other weak absortions becomes important. This makes the detection of this isotope a controversial fact. For instance, Reddy \etal\ (2002) did not find signatures of $^{6}$Li in HD82943, although they were using a blend of TiI in the Li region, instead of the SiI line used by \cite{Israelian03}. Other authors have also not found similar detections in other planet host stars (\cite{Mandell}, \cite{Ghezzi}). Uncertainties in the line lists can indeed lead to wrong determinations of $^{6}$Li abundance. Perhaps most importantly, the use of 3D models may change this panorama, since convection will produce an assymetry in the $^{7}$Li line similar to the one produced by the presence of  $^{6}$Li (\cite{Cayrel08}, \cite{Ghezzi}). This problem may affect the determination of $^{6}$Li abundances not only for metal-rich planet host stars but also for their metal poor
counterparts (see discussion in reviews by Asplund, Spite, Melendez, and Steffen).

In any case, present results suggest that signs of ``massive'' pollution are not generalized in planet host stars.

\section{$^{7}$Li}

The more common lithium isotope $^{7}$Li was produced during Big Bang nucleosynthesis and it can be produced in stellar interiors during AGB phase. It is depleted at a temperature of 2.5 million K primarily during the PMS in solar type stars but it can also be destroyed in stellar envelopes during MS if any mixing process exists. 

\begin{figure}[h!]
\begin{center}
 \includegraphics[width=10cm]{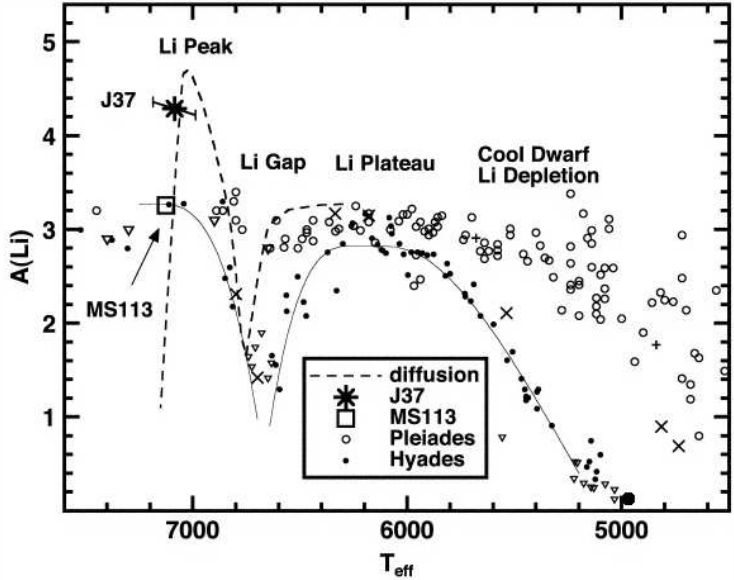} 
 \caption{\textit{Left panel:} Li abundances in the Hyades (detections: filled circles; upper limits: small inverted triangles; short-period binaries: mult crosses) and Pleiades (detections: open circles; upper limits: large inverted triangles; short-period binaries: plus signs) open clusters, in J37, and in MS113. From \cite{Deliyannis}. }
   \label{li7}
\end{center}
\end{figure}

The light element $^{7}$Li can also be used to trace pollution events. For instance, Deliyannis \etal\ (2002) discovered an extremely Li-rich dwarf, J37, a F-star in the open cluster NGC6633 (see Fig.\ref{li7}). Firstly they suggested upward, radiatively driven diffusion as the best explanation for this Li overabundance. However, later studies demonstrated a high Be abundance too (\cite{Ashwell}), a result that is in contradiction with radiative diffusion models. Furthermore, Laws \& Gonzalez (2003) showed that refractory elements were also overabundant, arriving at the conclusion that this star had accreted volatile-depleted material.

\begin{figure}[t!]
\begin{center}
 \includegraphics[width=10cm]{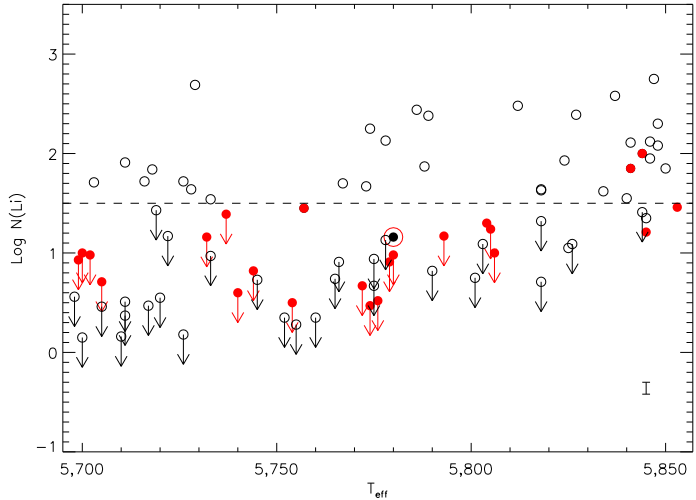} 
 \caption{Li abundances as a function of effective temperature for stars with (red filled circles) and without known planets (open circles), from \cite{Israelian09}.}
 \label{li7bis}
\end{center}
\end{figure}

A study of the differences of Li abundances in stars with and without planets was first carried out by \cite{King}. In that work they measured Li abundances for the binary system formed by 16CygA, a star without planets and a detectable Li abundance, and 16CygB, a planet-host that is Li depleted. This work was followed by several studies claiming that stars with planets have different Li abundances when compared to stars without detected companions (\cite{Cochran}; \cite{Gonzalez00}; \cite{Takeda}; \cite{Chen}; Israelian \etal\ 2004, 2009), though these results are not consensual (e.g. \cite{Ryan}; \cite{Luck}). A recent uniform study by  \cite{Israelian09} seems to confirm, however, that planet host stars have lower Li abundances in the solar temperature region (see Fig.\ref{li7bis}) and exclude metallicity, age, $v\,\sin{i}$, or activity as the cause of this anomaly (for more details see review by Israelian in this volume).

To explain the observed difference several possibilities exist. Pollution should be ruled out because it would have the opposite effect. On the other hand, it seems that stars with planets might have a different evolution. Extra mixing due to planet-star interaction, like migration, could take place (\cite{Castro}). The infall of planets might also affect the mixing processes of those stars (Theado \etal\, \textit{in prep.}) . Finally a long-lasting star-disc interaction during PMS may cause planet hosts to be slow rotators and develop a high degree of differential rotation between the radiative core and the convective envelope, also leading to enhanced lithium depletion (\cite{Bouvier}).

\section{Be}

\begin{figure}[t!]
\begin{center}
 \includegraphics[width=7cm]{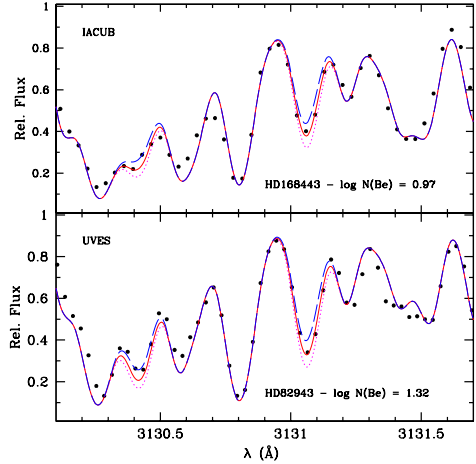} 
 \caption{Spectra in the BeII region (dots) for HD168443 and HD82943, and three spectral synthesis with different abundances, corresponding to the best fit (solid line) and to changes of $\pm$ 0.2 dex, respectively. From Santos et al. (2002). }
   \label{be_synth}
\end{center}
\end{figure}

Be is mainly produced by spallation reactions in the interstellar medium while it is burned in the hot stellar furnaces at a temperature of 3.5 million K. Li is depleted at much lower temperatures than Be. Thus, by measuring Li in stars where Be is not depleted (early G and late F) and Be in stars where Li is depleted (late G and K) we can obtain crucial information about the mixing, diffusion and angular momentum history of exoplanets hosts (\cite{Santos02}). Measurements of the abundance of this element are however difficulted by the fact that the only available lines are in the near-UV, a very blended region in metal-rich stars (Fig.\ref{be_synth}).

\begin{figure}[t!]
\begin{center}
 \includegraphics[width=10cm]{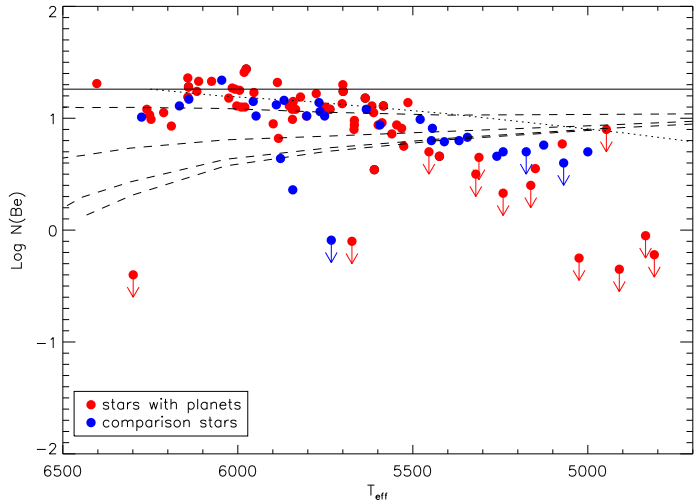} 
 \caption{Be abundances as a function of effective temperature for stars with (red circles) and without planets (blue circles). The dashed lines represent 4 Be depletion models of \cite{Pinsonneault} (Case A) with different initial angular momentum for solar metallicity and an age of 1.7 Gyr. The solid line represents the initial Be abundance of 1.26. The dotted line represents the Be depletion isochrone for 4.6 Gyr taken from the models including mixing by internal waves of \cite{Montalban}.}
   \label{be_teff}
\end{center}
\end{figure}

 There are not many works about Be abundances in planet host stars in the literature. Some of the first works made (\cite{Garcia-Lopez}; \cite{Deliyannis00}), had a very small number of objects. In Fig.\ref{be_teff} the largest samples made by the moment are plotted together, showing Be abundances as a funtion of effective temperature (Santos \etal\ 2002, 2004b; \cite{Galvez}; Delgado Mena \etal\, \textit{in prep.}). 
 
 At a first sight, there are not clear differences between stars with and without planets. Globally, Be abundances decreases from a maximum at 6100 K towards both higher and lower temperatures, in a similar way as Li abundandes behave. In the high temperature domain, the steep decrease with increasing temperature resembles the well known Be gap for F stars (Boesgaard \etal\ 1999). The decrease of the Be content towards lower temperatures is smoother and may show evidence for continuous Be burning during the main-sequence evolution of these stars. 
 
 In the temperature range where Li abundances are different in stars with and without planets, we do not observe any difference in Be abundances. This is something we can expect because for those temperatures, convective envelopes are not deep enough to bring the material of the convective envelope down to the layers where the temperature is high enough to burn Be. On the other hand, for the lowest temperatures of the range it seems that stars with planets have lower abundances when compared with stars for which no planet has been discovered. Unfortunately, the number of comparison stars in that temperature regime is still small to allow us to take a strong conclusion. 
 
In addition, in the low temperature region we find very low abundances of Be, in contradiction with models of Be depletion. In Fig.\ref{be_teff} , models with different initial rotation rates have been overplotted (Pinsonneault et al. 1990). As already noticed in Santos \etal\ (2004b,c), these models agree with the observations above roughly 5600 K, but while the observed Be abundance decreases towards lower temperatures when T$_{eff}$ $<$ 5600, these models predict either constant or increasing Be abundances. Even taking into account mixing by internal waves (\cite{Montalban}), Be depletion is still lower than observed. Although uncertainties in Be abundances for the coolest stars are large (significant systematic errors are not excluded), Be abundances are still overestimated by models. 
 
\section{Conclusions}
In this paper we reviewed the main results concerning the study of light element abundances in stars with planets. The main conclusions can be
listed as follows.
 
\begin{itemize}
 \item $^{6}$Li shows evidence that stars with planets may suffer (isolated) pollution events. However, the difficulty in deriving reliable abundances of this isotope make this a very debatable issue. An improvement of the line lists and in the determination of convective blue shifts is needed to understand how frequent are pollution events.
 \item $^{7}$Li has been found to be more depleted in stars with planetary companions in contrast with what happens in stars without detected planets. This trend is only observable in stars with temperatures in the solar range. This result suggests that some mechanisms is acting that is responsible for a higher Li depletion in planet-host stars. 
\item Be abundances in stars with an without planets do not follow the trend found for Li. No clear Be abundance differences seem to exist between the two groups of stars. More data is needed, specially for cooler stars, to understand the actual disagreements with the models for those temperatures.
\end{itemize}

\vspace{0.5truecm}

NCS would like to thank the support by the European Research Council/European Community under the FP7 through a Starting Grant, 
as well from Funda\c{c}\~ao para a Ci\^encia e a Tecnologia (FCT), Portugal, through a Ci\^encia\,2007 contract funded by FCT/M
CTES (Portugal) and POPH/FSE (EC), and in the form of grants reference PTDC/CTE-AST/098528/2008 from FCT/MCTES.

\end{document}